\documentclass[twocolumn,prb,amsmath,amssymb,amsfonts,floatfix,showpacs]{revtex4}

\usepackage{graphicx}% Include figure files
\usepackage{dcolumn}% Align table columns on decimal point
\usepackage{bm}% bold math
\usepackage{rotating} % rotate figure

\begin{document}

%\title{All-electron GW approximation self-energy based on the PAW method}
%\title{All-electron GW approximation based on the PAW method}
\title{
Electronic structure of superconducting gallium-doped germanium from ab-initio calculations.
}
\date{\today}
\author{S.~Leb\`egue}
\affiliation{
Laboratoire de Cristallographie, R\'esonance Magn\'etique et Mod\'elisations (CRM2, UMR CNRS 7036)
 Institut Jean Barriol, Nancy Universit\'e
 BP 239, Boulevard des Aiguillettes
 54506 Vandoeuvre-l\`es-Nancy,France
}

\begin{abstract}
	Using ab-initio calculations, we study the electronic structure of 
	gallium-doped germanium, which was found recently to be a superconductor, with a critical
	 temperature of $0.5$ Kelvins, and a particularly low density of Cooper pairs.
	 The calculations of the electronic properties reveal that no sign of an impurity band is observed,
	  and that the Fermi level lies in the valence band of Germanium. Moreover, the calculation of the phonon
	   frequencies shows that a new mode associated to the Ga atom is appearing, around $175$ cm$^{-1}$.
\end{abstract}

%                             % Classification Scheme.
%\preprint{APS/123-QED}

\maketitle
%%%%%%%%%%%%%%%%%%%%%%%%%%%%%%%%%%%%%%%%%%%%%%%%%%%%%%%%%%

The surprising discovery of the superconductivity of MgB$_2$\cite{Nagamatsu2001}, with a critical temperature
 of about $40$ Kelvins, has revived the interest of the scientific community for the systems presenting strong
 covalent bonds. Later on, the superconductivity phenomenon was observed in related systems like (among others)
 boron-doped diamond\cite{Ekimov2004}, silicon\cite{Bustarret2006}, and silicon-carbide\cite{Ren2007,Kriener2008}.
 These findings are explained by the fact that strong covalent bonds and light elements give rise to high phonon frequencies
  and large electron-phonon coupling, leading sometimes to superconductivity.
  In the framework of the standard electron-phonon coupling mechanism, ab-initio calculations are particularly useful, since
   it is possible to draw an almost complete picture of what is occurring, through the calculations of the electronic structure,
   the phonons spectra, and even an estimation of the critical temperature with the help of the McMillan formula\cite{CarbotteRMP}.
   Therefore, superconductivity has been well explained\cite{Blasereview} for example in the case of MgB$_2$\cite{An2001,Kortus2001}
    or boron doped diamond\cite{BlasePRL2004}.
    This is in contrast with the situation on copper oxides\cite{PickettRMP} or iron-based superconductors\cite{Kamihara2008},
    for which ab-initio calculations can reveal the electronic structure\cite{Lebegue2007,Singh2009,Yin2009},
    but are limited concerning the description of the mechanism leading to superconductivity.

      Very recently, a new member of the superconducting semiconductors has been found\cite{Herrmann2009}, by doping germanium with gallium atoms.
      The critical temperature was found to be $0.5$ K at ambient pressure, and a very low Cooper-pair density was observed.
      However, the information concerning the electronic structure of this material are still very limited. It is the aim
       of this publication to report results obtained by ab-initio calculations on Ga-doped germanium, in particular the
        electronic density of states, the bandstructure, and the phonon spectra.
   
To perform the calculations, we have used the code VASP (Vienna Ab-initio simulation package)\cite{vasp2,vasp},
 implementing the projector augmented waves method \cite{Bloechl} within density functional
theory (DFT) \cite{Hohenberg,Sham}. The Perdew Burke Ernzerhof\cite{gga} variant of the generalized gradient approximation (GGA)
 was used for the exchange-correlation (XC) potential.
We have built a germanium $2 \times 2 \times 2$ supercell (containing $16$ atoms),
 with one Ga atom substituting a germanium atom. This corresponds
  to a $6.25 \%$ doping, which is close to the percentage used in
   the experiments ($6 \%$)\cite{Herrmann2009}. Then, the system was allowed to relax completely.
   In this case, a $6 \times 6 \times 6 $ $k$-points\cite{Monkhorst} mesh was used,
    whereas to obtain the density of states and bandstructures, a denser $18 \times 18 \times 18 $ mesh
     was used. The plane wave cut-off was set to $500$ eV.

Due to the doping, the crystal structure is modified to accommodate the gallium atom in the cell.
In pure germanium, the Ge-Ge bond length is about $2.50~$ \AA~(when relaxed with PBE), while
 upon doping, the Ga-Ge distance is $2.47$ \AA, and the Ge-Ge distance is maintained to almost $2.50$ \AA.
 Therefore, the crystal structure is only slightly affected by the replacement of a Ge atom by a Ga atom, 
  and these structural changes remain localized around the Ga impurity.

\begin{figure}[h]
\includegraphics*[angle=0,width=0.38\textwidth]{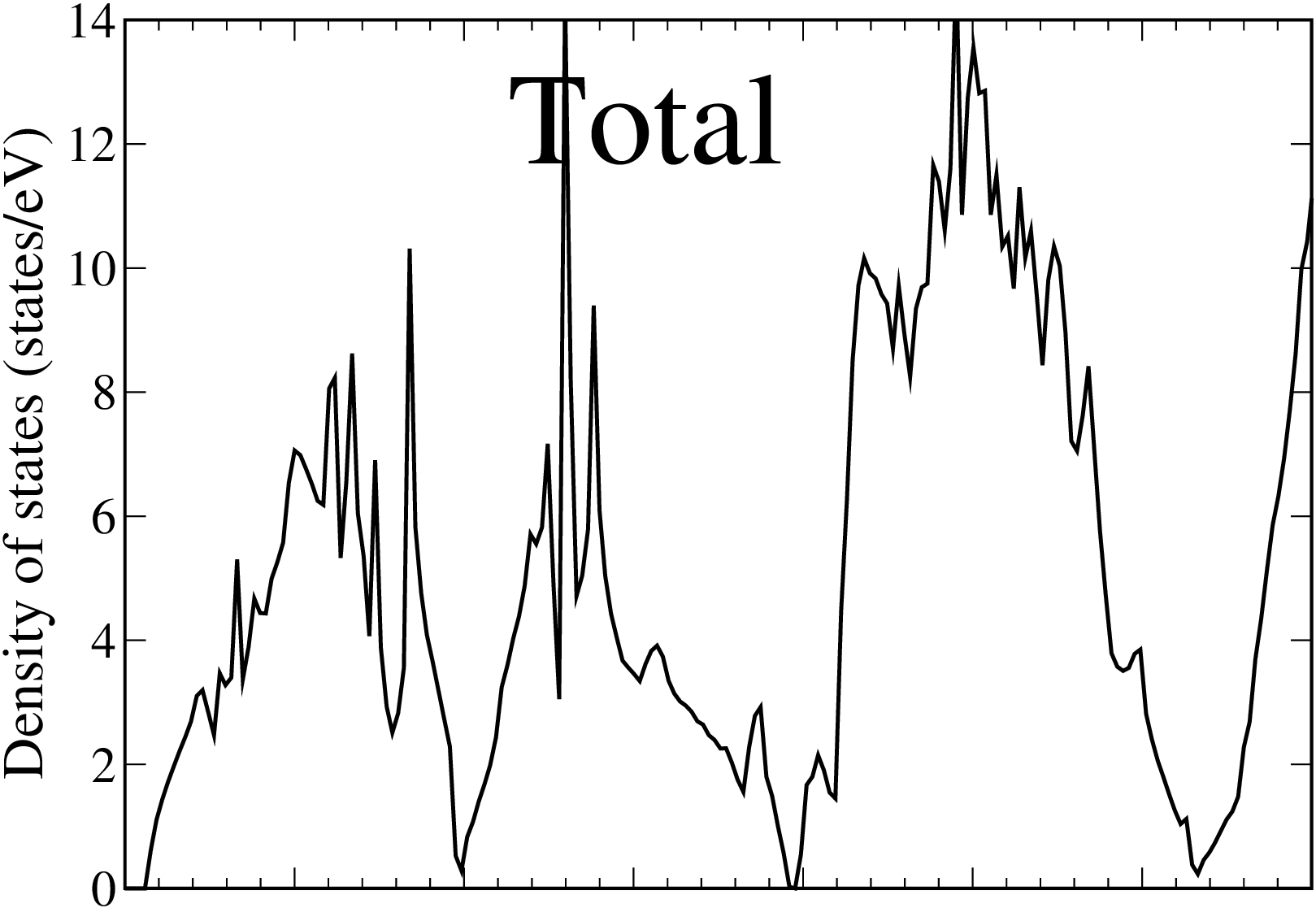}
\includegraphics*[angle=0,width=0.38\textwidth]{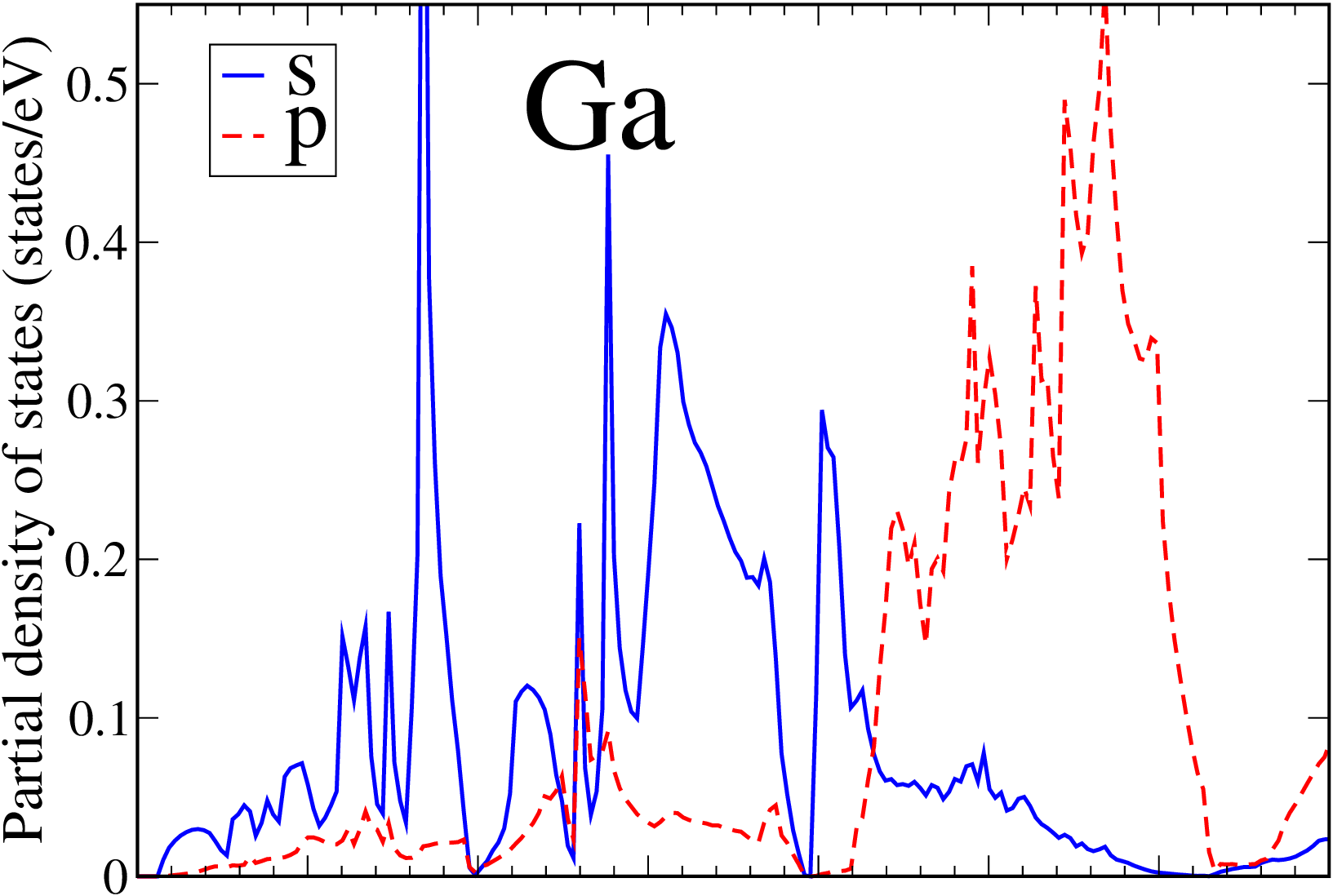}
\includegraphics*[angle=0,width=0.38\textwidth]{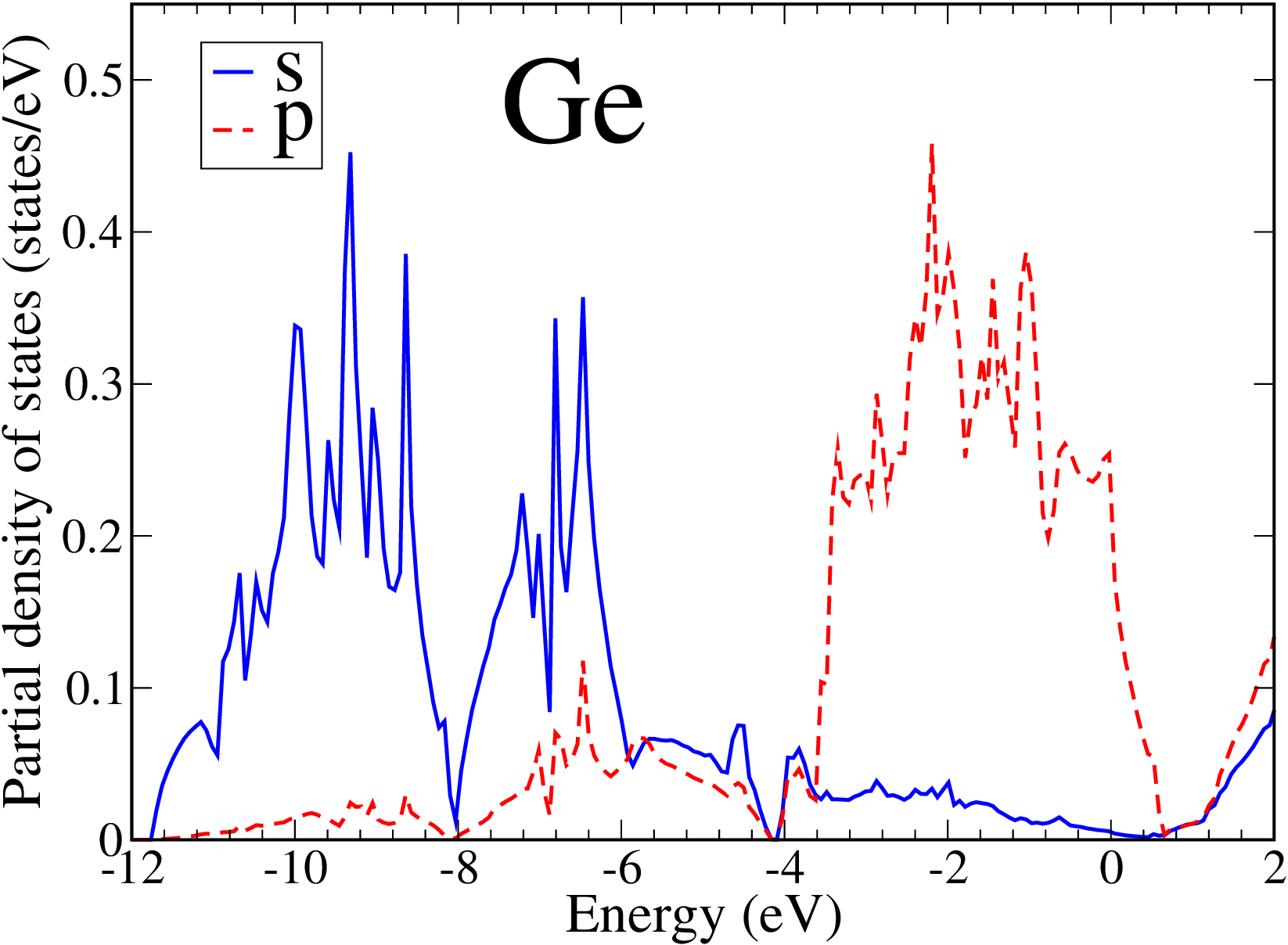}
\caption{
Total (upper plot) and partial (middle and lower plots) density of states.
The Fermi level is put at zero energy.
\label{fig:dos} 
}
\end{figure}

\begin{figure}[h]
\includegraphics*[angle=0,width=0.38\textwidth]{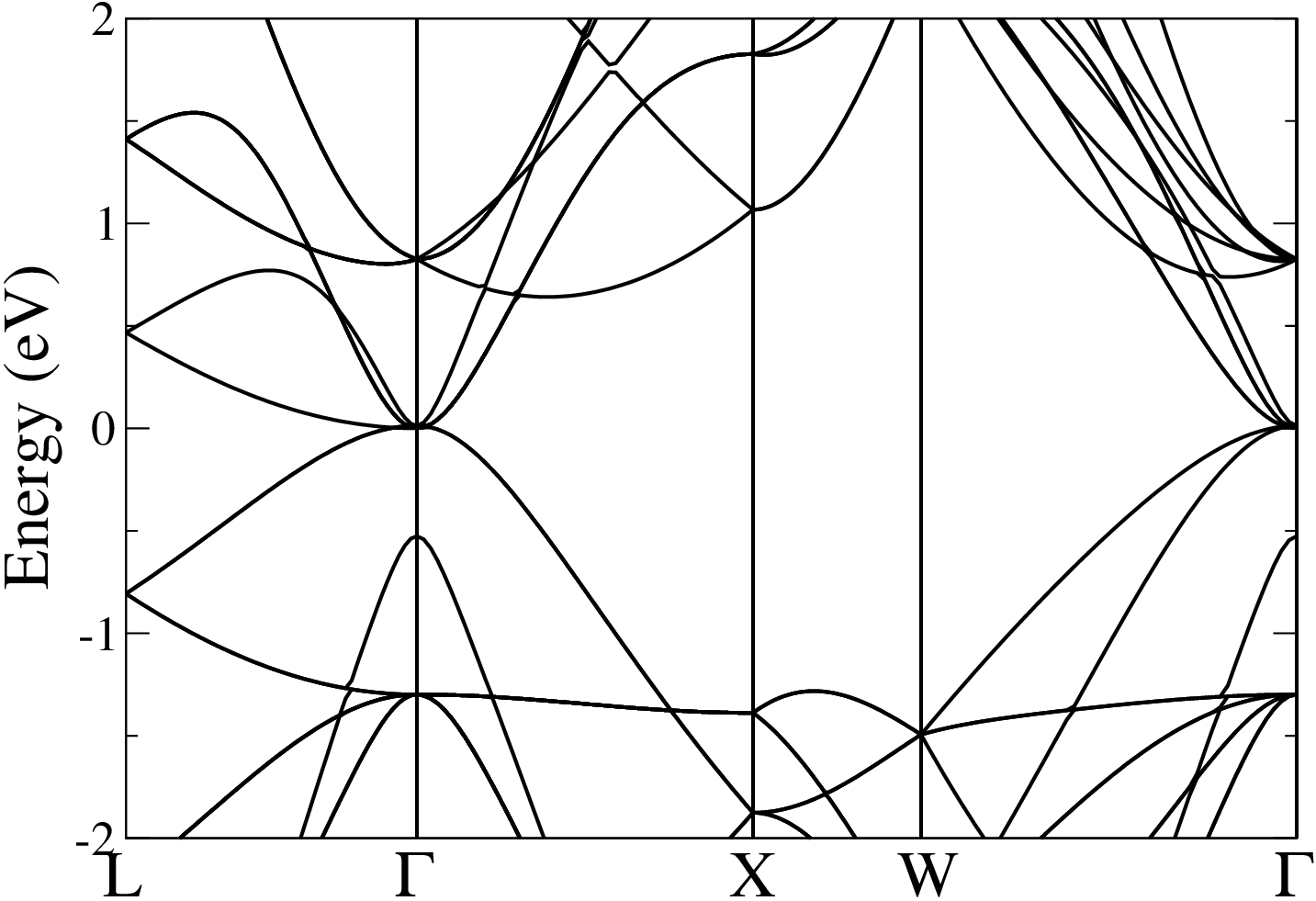}
\includegraphics*[angle=0,width=0.38\textwidth]{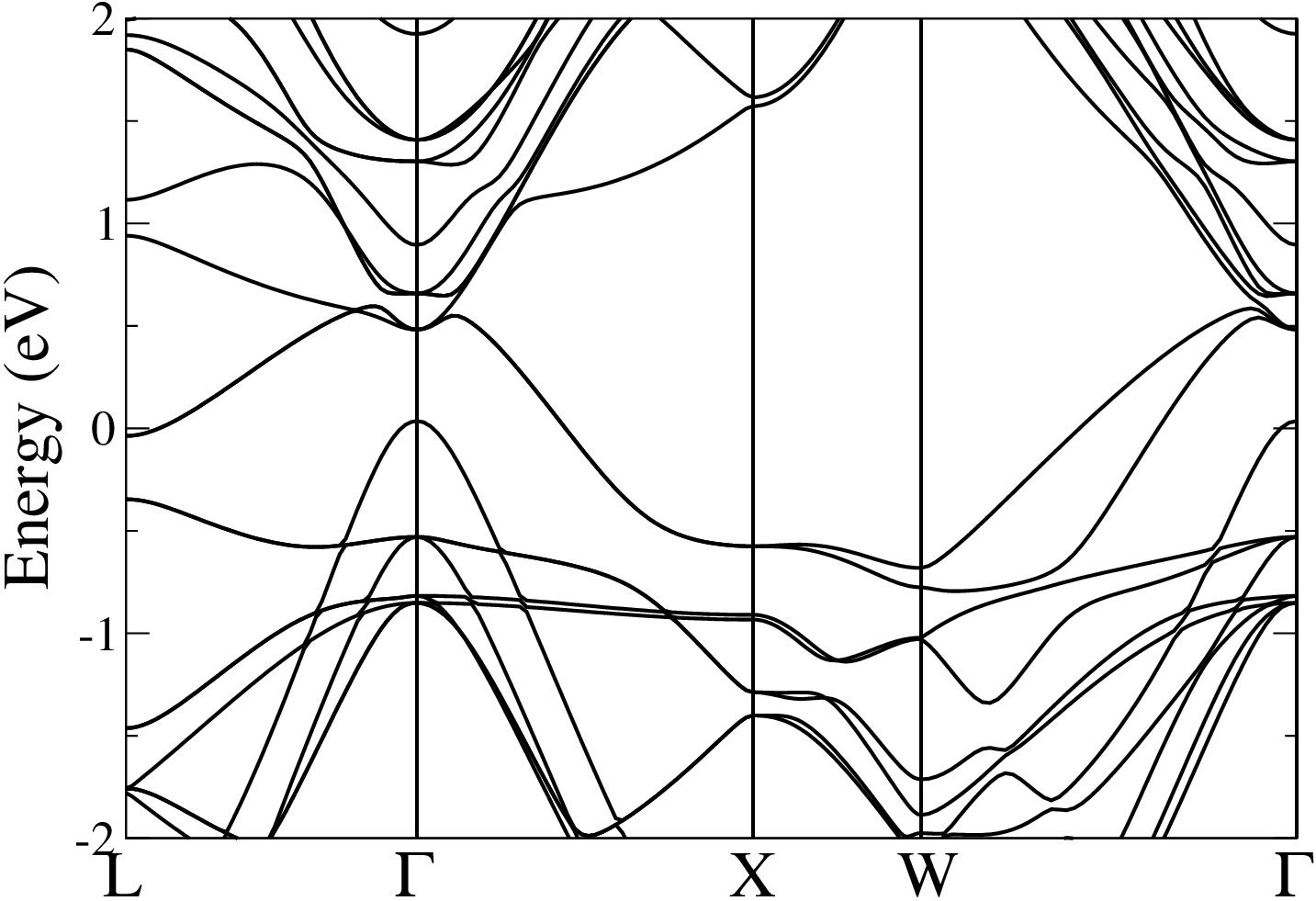}
\caption{
Bandstructures along high symmetry directions for pure (upper plot)
 and Ga-doped (lower plot) germanium, for a $2 \times 2 \times 2$ supercell.
 The Fermi level is put at zero energy. 
\label{fig:bands} 
}
\end{figure}

 In Fig. \ref{fig:dos}, we present our calculated total and partial density of states (DOS, PDOS)
  projected on the gallium atom and on one of the germanium atom neighbouring it. A metallic regime is clearly
   observed, with the Fermi level now being approximately $0.6$ eV below the valence band maximum of pure Ge (we are referring
    to a valence band maximum although here the density of states has always a finite value, due to the well known
     underestimation of the band gap with standard XC functional such as GGA). Also, a Fermi surface with three sheets (not shown
      here) emerges around the center of the Brillouin zone.
Moreover, there is no sign of an impurity band in the electronic gap, so at first sight
 the model of rigid bands doping seems likely, but this is not confirmed by the bandstructure
  calculations: in Fig. \ref{fig:bands}, the bandstructures of pure (upper plot) and 
  Ga-doped (lower plot) germanium are presented, for the $2 \times 2 \times 2$ supercell.
  Due to the doping, some degeneracies are left, the bandstructure is significantly modified, and therefore the picture of a simple rigid
   band model is not valid.

\begin{figure}[h]
\includegraphics*[angle=0,width=0.38\textwidth]{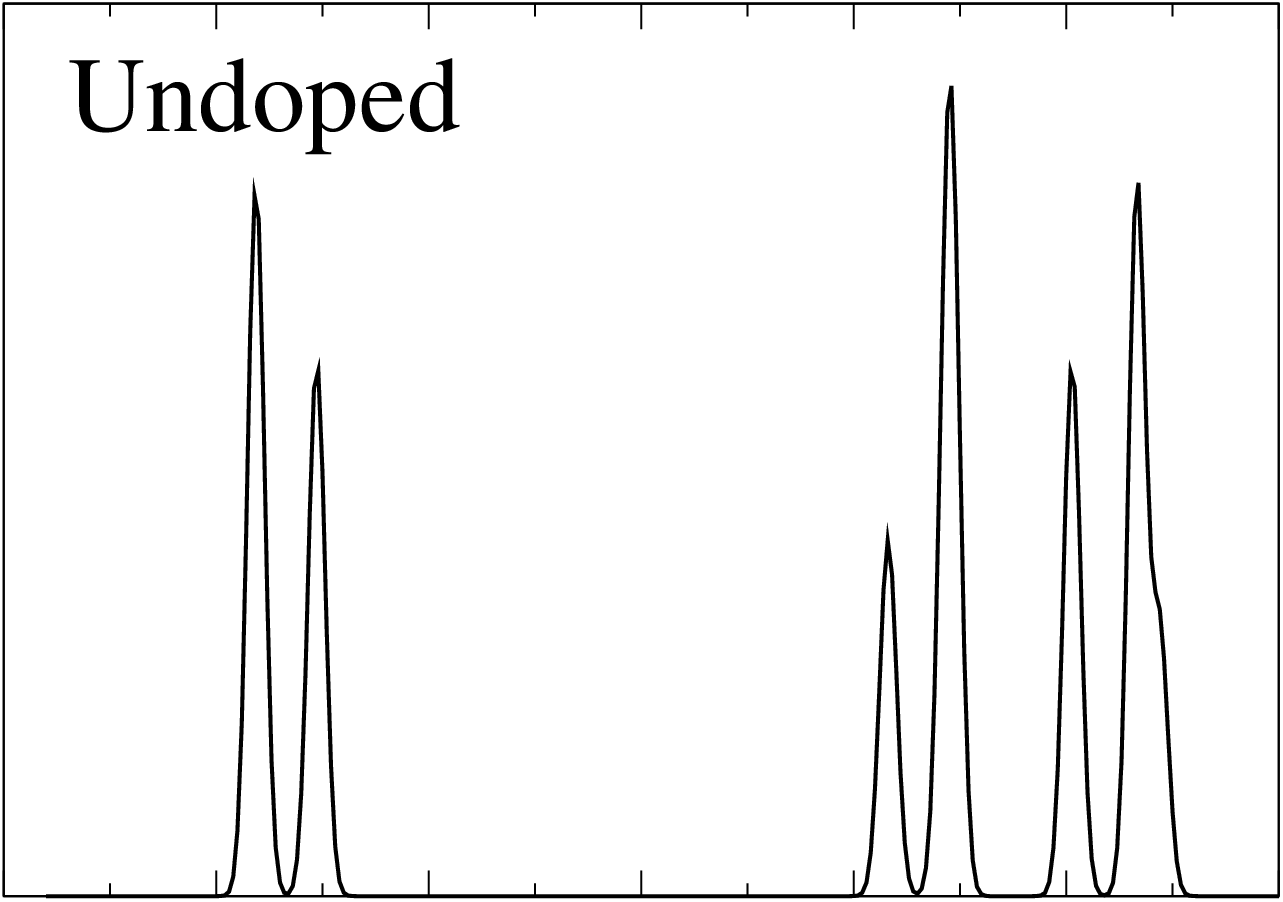}
\includegraphics*[angle=0,width=0.38\textwidth]{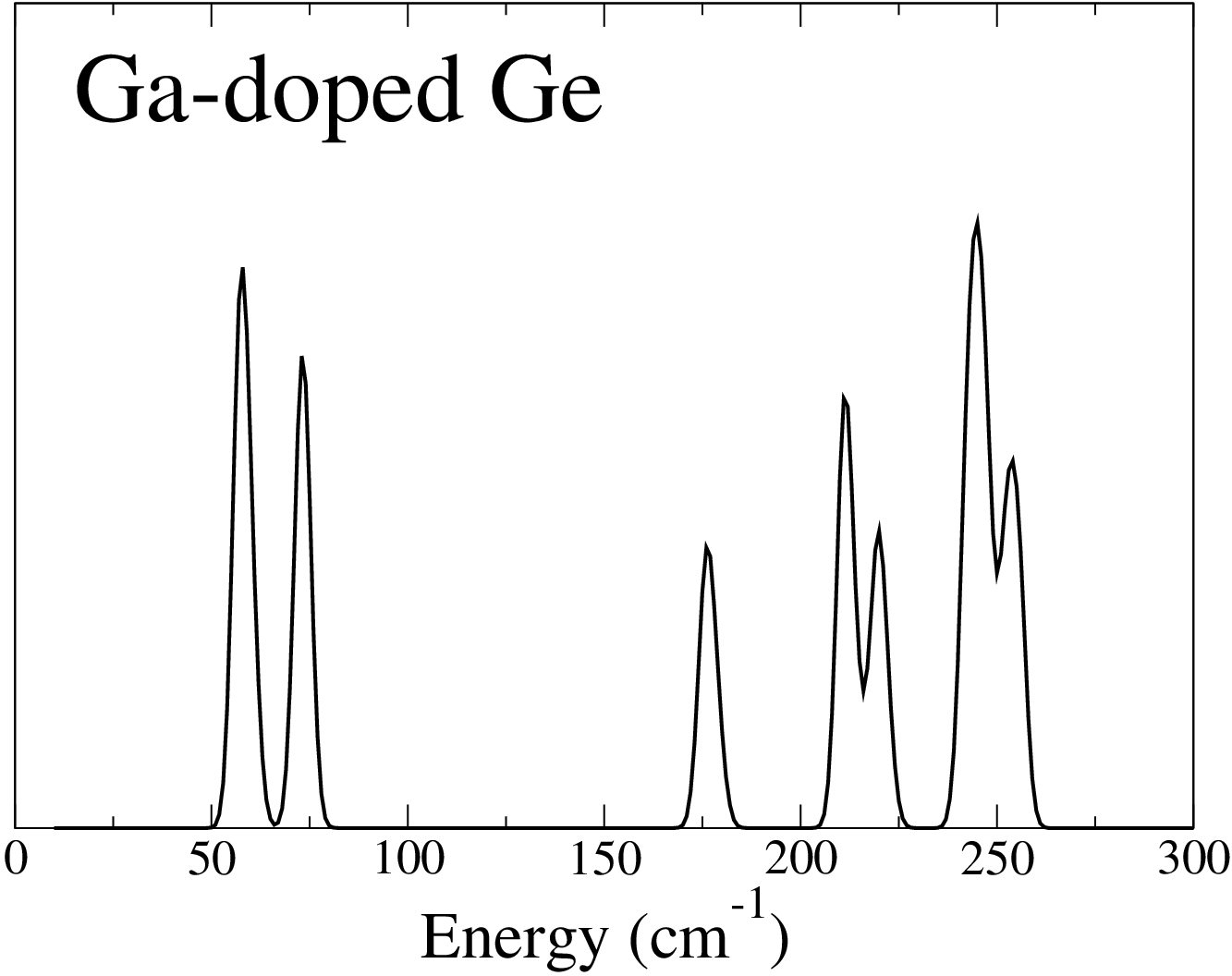}
\caption{
Phonons density of states (DOS) for pure germanium (upper plot) and
 Ga-doped germanium (lower plot). Energies are in cm$^{-1}$ and
  the y-axis magnitude is arbitrary.
\label{fig:dospho} 
}
\end{figure}

The phonons DOS ($\Gamma$ point only) of pure and Ga-doped Germanium are presented in Fig. \ref{fig:dospho}.
As expected, some degeneracies are lifted upon the introduction of the Ga atom, and the spectra is changed in
 comparison with the one of pure Ge.
However, this is only affecting the high energy part of the spectrum, with a softening of the frequencies. In particular, some modes are
  appearing around $175$ cm$^{-1}$ and $178$ cm$^{-1}$. By inspection of the eigenvectors, the new modes at 
$175$ cm$^{-1}$ are associated with a large displacement of the Ga atom. 
Therefore, it is possible that this specific mode is responsible of the superconductivity in Ga-doped Germanium, through a change of the
 electron-phonon coupling, as observed earlier for boron doped diamond\cite{BlasePRL2004}. However, this can be ensured only by the calculation
  of the Eliashberg function, since on the contrary, for B-doped SiC, it was found that the superconducting behaviour is controlled by
  the whole vibrational density of states\cite{Noffsinger}.

In summary, we have computed the electronic and vibrational properties of Ga-doped Germanium.
We observed that the effect of doping is to shift the Fermi level in the valence band of Germanium.
Concerning the vibrational properties, a mode around $175$ cm$^{-1}$ is associated with Ga atoms.
We hope that our work will stimulate further theoretical work, such as the calculation of the strength
 of the electron-phonon coupling and an estimate of the superconducting transition temperature, 
  or further experiments, such as photoemission or neutron diffraction experiments, to obtain an even better\
  picture of the properties of Ga-doped Germanium.

I acknowledge financial support from ANR grant No. ANR-06-NANO-053-02 and ANR Grant ANR-BLAN07-1-186138 as well as
 CINES-CCRT for computer time.

%%%%%%%%%%%%%%%%%%%%%%%%%%%%%%%%%%%%%%%%%%%%%%%%%%%%%%%%%%%%%%%%%%%%%%%
%                         BIBLIOGRAPHY
%%%%%%%%%%%%%%%%%%%%%%%%%%%%%%%%%%%%%%%%%%%%%%%%%%%%%%%%%%%%%%%%%%%%%%%
%\bibliography{sup}

\end{document}